\documentclass[runningheads]{svmult}

\usepackage{makeidx}   
\usepackage{graphicx}  
\usepackage{subeqnar}  
\usepackage{multicol}  
\usepackage{physprbb}  
\makeindex             

\newcommand{\cent}[1] {\begin{center}#1\end{center}}
\newcommand{\doublint}{\int\rule{-3.5mm}{0mm}\int} 
 
\newcommand{\vecbm}[1]{\mbox{\boldmath$#1$}}

\newcommand{\lra} {$\leftrightarrow$}
\newcommand{\vecb}[1]{\mbox{\bf#1}} 
\newcommand{\lora}{{\boldmath$\longrightarrow$}}

   \newcommand{\mlora} {\bf\longrightarrow}
\renewcommand{\l}{\left}
\renewcommand{\r}{\right}
\usepackage{graphicx}
\usepackage{amsmath,amssymb,amsfonts}
\begin{document}
\title*{Thermo-Statistics or Topology\\of the Microcanonical Entropy
  Surface}
\toctitle{Thermo-Statistics or Microcanonical Topology}
%
%
\titlerunning{Thermo-Statistics or Microcanonical Topology}
%
\author{D.H.E. Gross\inst{1}\inst{2}}
\authorrunning{D.H.E. Gross}
%
%
\institute{Hahn-Meitner Institute
\and Freie Universit{\"a}t Berlin,\\
Fachbereich Physik.\\
Glienickerstr. 100\\
14109 Berlin, Germany}

\maketitle              

\begin{abstract}
  Boltzmann's principle $S(E,N,V\cdots)=\ln W(E,N,V\cdots)$ allows the
  interpretation of Statistical Mechanics of a {\em closed} system as
  Pseudo-Riemannian geometry in the space of the conserved parameters
  $E,N,V\cdots$ (the conserved mechanical parameters in the language
  of Ruppeiner~\cite{ruppeiner95}) without invoking the thermodynamic
  limit.  The topology is controlled by the curvature of
  $S(E,N,V\cdots)$. The most interesting region is the region of
  (wrong) positive maximum curvature, the region of phase-separation.
  This is demonstrated among others for the {\em equilibrium} of a
  typical {\em non-extensive} system, a self-gravitating and rotating
  cloud in a spherical container at various energies and
  angular-momenta. A rich variety of realistic configurations, as
  single stars, multi-star systems, rings and finally gas, are
  obtained as {\em equilibrium} microcanonical phases.  The global
  phase diagram, the topology of the curvature, as function of energy
  and angular-momentum is presented. No exotic form of thermodynamics
  like Tsallis~\cite{tsallis99,tsallis88} non-extensive one is
  necessary.  It is further shown that a finite (even mesoscopic)
  system approaches equilibrium with a change of its entropy $\Delta
  S\ge 0$ (Second Law) even when its Poincarr\'e recurrence time is
  not large.
\end{abstract}
\section{Introduction}
Why this paper?\\
Since more than $100$ years does thermo-statistics emphasize the
canonical- or the grand-canonical ensemble in the thermodynamic limit
\index{thermodynamic limit} as the appropriate microscopic description
of an equilibrized system. Here a {\em homogeneous}, practically
infinite, system is controlled by intensive parameters like the
temperature.  Though in textbooks the microcanonical ensemble is
considered as the fundamental ensemble, due to mathematical
difficulties it is quickly left in favor of the canonical ones.
Intensive variables like temperature, pressure and chemical potential
are used instead of the mechanical defined extensive energy, volume
and particle number.

The intensive variables are even canonized to found the axiomatic
definition of an orthode~\cite{gallavotti99}\index{orthode} from where
Statistical Mechanics (at least its intensive or canonical form) is
deduced. Lebowitz~\cite{lebowitz99a,lebowitz99} considers the
thermodynamic limit \index{thermodynamic limit} and Elliott
Lieb~\cite{lieb98a} extensivity\index{extensivity}, which also needs
the thermodynamic limit to ignore surface effects, as the condition
sine-qua-non.

In the thermodynamic limit $\lim_{V,N\to\infty,~~N/V=\rho}$
\index{thermodynamic limit} \index{volume vs. surface} all surface
effects may be ignored.  --~Usually~-- It is clear that this approach
cannot do justice to phase separations\index{phase-separation}. In
fact, the gain in entropy when a system splits into different phases
by interphase surfaces\index{interphase surfaces} 
\index{inhomogeneity at equilibrium} is the essence of phase
transitions of first order.  A liquid-gas transition 
\index{liquid-gas transition} is experimentally detected just by the
interface between the liquid and the gas.  Consequently, in the
(grand)-canonical approach, phase-transitions are indicated by the
Yang-Lee singularities~\cite{yang52} where the grand-canonical
potentials are non-analytic in $z=e^{\beta\mu}$ or singular. These
indicate the break-down of the (grand)-canonical formalism. In
remarkable contrast, the microcanonical ensemble has no problems at
phase-separations and the microcanonical density of states remains
single-valued and multiply differentiable in all conserved control
parameters also here, see below. This is certainly the strongest
argument in favor of the fundamental significance of the
micro-ensemble.

At phase-separation the entropy $S(E,N,\cdots)$ has a positive
curvature. Ruppeiner's Riemannian geometry of
fluctuations~\cite{ruppeiner95} must be reformulated there as
Pseudo-Riemannian\index{Pseudo-Riemannian geometry}. This leads to a
negative heat capacity and a violation of Clausius' formulation of the
Second Law (``heat flows always from the hot to the cold system'').
\index{Clausius formulation of Second Law violated: 
Heat can flow from cold to hot}
Phase-separations demand an essential, fundamental, change of
conventional classical thermo-statistics.  Thermodynamics, however,
was invented in the $19$.century to explain the working of steam
engines\index{steam engines}. I.e. one of its primary issues were
just phase-separations.

What was said applies to large systems with short-range forces.  The
largest systems in nature, self-gravitating astro-physical systems,
are subjected to forces (gravitation)\index{gravitation} with a range
comparable to the linear extension of the system. These systems are
naturally inhomogeneous and non-extensive\index{non-extensivity}. A
description of their equilibrium by intensive variables with a
homogeneous spatial distribution misses these most interesting
situations. Ironically, \index{thermodynamic limit} the thermodynamic
limit does not apply to these really large systems.
Tsallis~\cite{tsallis99}\index{Tsallis} on the other hand addresses
non-extensive systems explicitely but keeps the description in terms
of intensive variables which fix the relevant conserved parameters
only on average. However, non-extensive systems are usually not
self-averaging. He believes the equilibrium statistics of Hamiltonians
systems demands a new definition of entropy. This, is not so~\cite{gross175}, 
see section~\ref{gravit}.

It is well known that self-gravitating systems\index{gravitation}
collapse to a star in a more or less void background at low energies,
the ``gravothermal catastrophe''\index{gravothermal catastrophe}
\cite{lyndenbell68,thirring70}.  This is of course quite interesting
but outside of any homogeneous canonical thermodynamics. There is
nothing peculiar with this in microcanonical thermodynamics.
Certainly, angular-momentum\index{angular momentum and gravity} can
change this essentially.  However, the microcanonical equilibrium
configuration of a self-gravitating system under larger
angular-momentum has not been investigated yet
(exception~\cite{gross187}). I will show in section~\ref{gravit} how
the competition between gravitational collapse\index{gravitational
  collapse} and centrifugal disruption\index{centrifugal disruption}
leads in a natural manner to a breaking of rotational symmetry and to
multi-star configurations with a large variety of different but quite
realistic configurations. In this section the global phase diagram of
self-gravitating and rotating many-body systems as function of energy
and angular-momentum is presented, c.f.~\cite{gross187}. It is for the
first time that these various realistic stellar configurations are
interpreted as global equilibrium configurations.

The thermodynamic limit\index{thermodynamic limit} is also invoked
since Boltzmann to deduce the Second Law\index{Second Law} of
Thermodynamics~\cite{lebowitz99,gaspard97,lavanda90,lieb02} from
microscopic reversible dynamics.  Then Zermelo's~\cite{zermelo97}
paradox\index{Zermelo's paradox} becomes blunted as the Poincarr\'e
recurrence time $t_{rec}$ \index{Poincarr\'e recurrence time} is much
longer than any physically relevant observation time. This is
different for a finite, eventually small system.


By all these reasons a reinvestigation of the microscopic foundation
of Statistical Mechanics starting from Newtonian {\em reversible}
mechanics of the {\em finite} many-body system using a minimum of
assumptions but avoiding the thermodynamic limit
\index{thermodynamic limit} is highly needed. The ``Geometric
Foundation of Thermo-Statistics'' proposed in~\cite{gross186} and
further developed here offers a new, deeper, and much simpler
understanding of the microscopic foundation of Thermodynamics.
\section{The few essentials of Statistical Mechanics}
\subsection{Why does conventional statistics like the thermodynamic 
  limit?}  

The relative fluctuations of a macroscopic observable $A$ in pure
phases of an extensive system vanish in the large $N$ limit:
\begin{gather*}
\frac{<A^2>-<A>^2}{<A>^2}\propto \frac{1}{N}
\end{gather*}
We call this behavior\index{self-averaging} {\em self-averaging}. Then
the probability aspect of statistics becomes unimportant.

Here we want to study also non-extensive situations, therefore, we are
not allowed to go to the thermodynamic limit. Fluctuations must be
taken seriously.
\subsection{Obsolete gospels of conventional thermo-statistics}
Then, most axioms which are mistaken to be fundamental for Statistical
Mechanics\index{unnecessary gospels of thermo-statistics} even turn
out to be violated:\index{unnecessary gospels of thermo-statistics}
\begin{itemize}
\item Phase transitions ({\em do not}) exist only in the thermodynamic limit
\item Specific heat is ({\em not}) $c\sim <\!\!(\delta E)^2\!\!>\;\; >0$
or $d T/d E > 0$
\item Heat does ({\em not}) always flow from hot to cold
\item Thermodynamic stability does ({\em not}) necessarily imply the
  concavity of\\ $S(E,N,\cdots)$
\item ({\em No }) extensivity of $S$, ({\em no }) scaling with $N$
\item ({\em No}) unique Legendre mapping, $T{\mlora} E$, etc. 
\item Rise of entropy is ({\em not}) necessarily connected to trend
  towards uniformization
\item Second Law ({\em not})\index{Second Law} only in infinite
  systems, i.e. second law ({\em not}) only when the recurrence time
  is much larger than the observation-time $t_{rec}\gg t_{obs}$
\item A system at equilibrium is ({\em not}) necessarily an orthode
\index{orthode} in the sense of Gallavotti~\cite{gallavotti99}.
``Boltzmann's heat theorem''\index{heat theorem}, i.e.
\begin{equation}
\frac{dE(T,P)}{T}+\frac{PdV(T,P)}{T}
\end{equation}
is ({\em not}) necessarily a total differential $dS(T,P)$,
 because $S(T,P)$ is not always a smooth, one valued,
 function of $(T,P)$, see section \ref{order parameter}, order parameter.
\end{itemize}
Violations of these gospels seem shocking statements:\\
Lebowitz~\cite{lebowitz99a} and Lieb\cite{lieb97,lavanda90} believe
these make thermo-statistics impossible to exist. Nevertheless these
violations are valid building stones of statistics.  They are even
necessary for thermo-statistics to apply to the original goal of
Thermodynamics, the description of phase-separation, the scenario in
which steam engines\index{steam engines}\index{phase-separation} do
work.  At closer inspection these violations are not so strange.
Recapitulating the main roots of statistical mechanics we will see
that it makes a lot of sense to formulate it without invoking the
above axioms and without using the thermodynamic limit.
\index{thermodynamic limit} The only essential axiom needed to define
equilibrium statistics is Boltzmann's principle c.f. section
\ref{boltzmanns principle}, eq.(\ref{boltzmentr1}), once we agree not
to use the thermodynamic limit. To concentrate on this is a great
advantage as this principle has a very simple {\em geometrical}
meaning.

It is a benefit of the new, extended theory which I offer here, that
by reducing its axiomatic basis to this single principle it applies
also to Hamiltonian non-extensive systems\index{non-extensivity} and
among them to the really large systems as astrophysical ones, which
are far larger than the thermodynamic ``limit'' of any homogeneous
system allows. A whole new world for applications of thermo-statistics
opens.  Of course, then one cannot separate volume from surface
effects.\index{volume vs. surface} This is anyhow dubious for
non-extensive systems or at phase-separation\index{phase-separation}.

Here however, I must make it very clear that in any cases where the
thermodynamic limit \index{thermodynamic limit} of a homogeneous phase
does exist, the geometrical theory is in complete agreement with
conventional Thermodynamics and conventional extensive Statistical
Mechanics.

\subsection{Thermodynamics, a probabilistic theory; Control parameters}
\index{macroscopic parameters}
\index{redundancy of Thermodynamics} Thermodynamics is a
macroscopic description of a many-body system within a few ($M\sim 3$)
macroscopic control parameters and where ($6N-M\gg M$) dof's remain
uncontrolled.  Therefore, Thermodynamics describes all systems with
the same $M$ simultaneously. All systems under the same macroscopic
constraints are simultaneously addressed by the theory.  Statistical
mechanics describes the whole $6N-M$ dimensional manifold, i.e. all
points in then $N$-body phase-space with same energy $E$ , the
microcanonical ensemble $\cal{E}$\footnote{We denote manifolds in
  phase-space by calligraphic letters like $\cal{E}$.}. Consequently,
it gives only {\em probabilistic}
\index{thermo-statistics, a probabilistic theory} predictions how the
{\em average} of the systems in the ensemble behave~\cite{gross183}.

A large extensive system in a pure phase is
self-averaging\index{self-averaging}. In the thermodynamic
limit \index{thermodynamic limit} a sharp peak of the probability
distribution guarantees the identity of the average with the most
likely configuration.

Bur what if the thermodynamic limit does not exist like for a
non-extensive system? For a small system like a nucleus or an atomic
cluster the same measurement must be performed very often and the
average be taken before its thermodynamic behavior can be seen.
\subsection{Boltzmann's principle, the microcanonical ensemble 
\label{boltzmanns principle}}
The key quantity of statistics and thermodynamics is the {\em entropy}
$S$\index{entropy}.  Its most fundamental definition is as the
logarithm of the {\em area} $W(E)$ of the manifold $\cal{E}$ in the
N-body phase-space by Boltzmann's principle~\cite{einstein05d}:

\begin{eqnarray} 
W(E,N,V)&=&\epsilon_0 tr\delta(E-H_N)\nonumber\\
tr\delta(E-H_N)&=&\int{\frac{d^{3N}p\;d^{3N}q}{N!(2\pi\hbar)^{3N}}
\delta(E-H_N)}.\label{wenv}
\end{eqnarray}
\begin{equation}
\fbox{\fbox{\vecbm{$S=k$\cdot$lnW$}}}\label{boltzmentr1}\end{equation}
\index{area of microcanonical manifold}\index{Boltzmann's principle}
\index{entropy}\index{microcanonical entropy}\\
{\em Boltzmann's principle is the only axiom necessary for thermo-statistics.}
With it Statistical Mechanics and also Thermodynamics become
{\em geometric} theories. For instance all kinds of phase-transitions
are entirely determined by {\em topological} peculiarities of
${\cal{E}}(E,N,\cdots)$ and thus of $S(E,N,\cdots)$ see below.
\subsubsection{Local or global constraints?}~\\
\index{locality in macroscopic conserved observables} In
microcanonical statistics the ``extensive'', better conserved, control
parameters as energy, volume, particle number etc. can be determined
macroscopically sharp. There may well be small, microscopic violations
of some microscopic conservation laws due to the non-ideal nature of
the container. Therefore, we allow small fluctuations in these
microscopically conserved quantities. Evidently, it does not matter
whether the entropy $S(E,N,V)$ has an extremum or not. Its {\em local}
value is significant. It is uniquely defined by eqs.
(\ref{wenv},\ref{boltzmentr1}) as a high-dimensional integral. It is
thus everywhere multiply differentiable, -- certainly the most
important difference to canonical statistics. This is especially
important at phase-separations where the curvature of $S(E,N,V)$ is
positive c.f.  section~\ref{phase-separation}.
\subsubsection{Why not canonical?}~\\
In the canonical statistics, also in Tsallis ``non-extensive
statistics''\index{Tsallis} e.g. the energy is fixed in the mean by
Lagrange parameters like $\beta=1/T$. This works only if the
microcanonical ensemble is self-averaging\index{self-averaging}. Now
for non-extensive situations like at phase-separations e.g. the energy
$E(T)$ as function of the Lagrange parameter like $T$ is multi-valued
c.f.  section~\ref{ambiguity}(ambiguity $\cdots$).
\index{multi-valuedness of entropy in terms of intensive variables}
This leads to the (in view of standard Thermodynamics) surprising
negative heat-capacity c.f. section~\ref{negative
  heat-capacity}(negative heat capacity $\cdots$),
\index{negative heat capacity} which is well documented even
experimentally
c.f.\cite{thirring70,gross95,chbihi95,lyndenbell95a,gross158,gross150,lyndenbell99,gross171,gulminelli99a,casetti99a,schmidt00,dAgostino00,schmidt01,ispolatov01}.
\footnote{In fact the paper \cite{gross158} pointed explicitely to the
  fundamental failure of the canonical ensembles near first order
  phase transitions in general and its non-equivalence to the
  fundamental micro-ensemble which shows a negative heat capacity
  there. It was rejected by Gary S. Grest, the Divisional Associate
  Editor of statistical mechanics of PRL, June 3 1997, with the
  argument: ``I am not convinced that the microcanonical ensemble is
  more physical than the canonical ensemble. After all, phase
  transitions in the experimental world are at constant temperature
  and not at constant energy.''  If this would be true, a ship would
  not have been able to sail on the surface of the ocean, America
  would never have been discovered and PRL would not even exist.}
\subsubsection{At points of negative curvature of $S(E,N,\cdots)$ the 
  canonical ensemble is not an orthode}~\\ \index{orthode} In chapter
1.5 of his book\cite{gallavotti99} Gallavotti presented an axiomatic
deduction of thermo-statistics starting from the definition of an
orthode.

Following Boltzmann's heat theorem~\cite{boltzmann1896} he defines an
ensemble to be an orthode when an infinitesimal change of the energy
$dE$ and volume $dV$ makes
\begin{equation}
\frac{dE}{T}+\frac{PdV}{T}
\end{equation}
an exact differential, at least in the thermodynamic limit. Here $T$
the ``temperature'' is the average kinetic energy per particle and $P$
the ``pressure'' is defined as the average momentum transfer per unit
time and unit surface area of the container.\index{heat theorem} 

Clearly this definition is of little help for a non-extensive system
and/or when the thermodynamic limit does not exist, where a given $T$
or $P$ does not fix the energy or volume. It fails in situations where
the canonical ensemble is not equivalent to the microcanonical one,
i.e. also at phase-transitions.
\section{Equilibrium Thermodynamics}
\index{classical equilibrium}
\subsection{Phase transitions}\label{phase-separation}
\subsubsection{Relation to Yang-Lee theory}~\\
In conventional extensive thermodynamics phase transitions are
indicated by the Yang-Lee zeros of the grand-canonical partition sum
$Z$ {\em in the thermodynamic limit}. In order to generalize the
definition of phase transitions also to non-extensive
systems\index{non-extensivity} I start for the moment with the Laplace
transform from the microcanonical density of states $e^S$ to the
grand-canonical one (here the discreteness of $N$ does not matter):
\begin{eqnarray}
 Z(T,\mu,V)&=&\doublint_0^{\infty}{\frac{dE}{\epsilon_0}\;dN\;e^{-[E-\mu
N-TS(E)]/T}}\\
&=:&\frac{V^2}{\epsilon_0}\doublint_0^{\infty}{de\;dn\;e^{-V[e-\mu
n-Ts(e,n)]/T}}\label{grandsum}\\
&\approx&\hspace{2cm}e^{\mbox{ cons.+lin.+quadr.}}\nonumber
\end{eqnarray}
The linear term is explicitely put to $0$ by solving
\begin{eqnarray}
\frac{1}{T}&=&\left.\frac{\partial S}{\partial E}\right|_{e_s,n_s}
\label{statpoint1}\\
\frac{\mu}{T}&=&-\left.\frac{\partial S}{\partial N}\right|_{e_s,n_s}
\label{statpoint2}\\
\left(\frac{P}{T}\right.&=&\left.\left.\frac{\partial S}
{\partial V}\right|_{e_s,n_s}\right).\label{statpoint3}
\end{eqnarray}
If $s(e,n)$ is concave (has negative principal curvatures), and there
is a single solution $e_s$,$n_s$ of eqs.~(\ref{statpoint1})
and~(\ref{statpoint2}), the stationary point, where
integral~(\ref{grandsum}) is a double Gaussian integral along the two
principal curvatures $\vecbm{v}_1,\vecbm{v}_2$ and the free-energy
density is:
\begin{displaymath}
\frac{F(T,\mu,V)}{V}=\frac{-T\ln(Z)+\mu N_s}{V}\to e_s-Ts_s
+\frac{T\ln{(\sqrt{(-\lambda_1)}\sqrt{(-\lambda_2)})}}{V}
+o(\frac{\ln{V}}{V})
\end{displaymath}

The curvature matrix $c(e,n)$ of $s(e,n)$
\begin{eqnarray}
c(e,n)&=&\left(\begin{array}{rcl}
\frac{\partial^2 s}{\partial e^2}& \frac{\partial^2 s}{\partial n\partial e}\\
\frac{\partial^2 s}{\partial e\partial n}& \frac{\partial^2 s}{\partial n^2}
\end{array}\right)\label{curvmatr}\\
\mbox{has the  eigenvalues :}
\lambda_1\ge\lambda_2&&\hspace{1cm}\mbox{\lora eigenvectors :}\hspace{1cm}
\vecbm{v}_1,\vecbm{v}_2\\
\mbox{ Hessian }d(e,n)&=&\left\|c(e,n)\right\|=\lambda_1\lambda_2 
\label{curvdet}
\end{eqnarray}
In general $\lambda_1$\index{curvature, largest eigenvalue} can have
either sign. This leads to a new, deeper definition of thermodynamic
phases:
\subsubsection{Classification by the local topology of curvature}
\begin{itemize}
\item A single stable phase is defined by $\lambda_1<0$. Here $s(e,n)$
  is concave (downwards bended) in both directions,
  fig.~(\ref{concave}). There is only a single solution of:
\begin{eqnarray}
\frac{1}{T}&=&\left.\frac{\partial S}{\partial E}\right|_{e_s,n_s}\\
\frac{\mu}{T}&=&-\left.\frac{\partial S}{\partial N}\right|_{e_s,n_s}.
\end{eqnarray}
Here is a one to one mapping of the (grand)-canonical \lra the
micro-ensemble. The (grand)-canonical has in the thermodynamic
limit \index{thermodynamic limit} the same analytical properties 
as the micro-ensemble. It is everywhere smooth, multiply
differentiable.  There are no discontinuities neither in $\ln{Z}$ nor
its derivatives.
\begin{figure}[h]
\cent{\includegraphics*[bb =  53 14 430 622,
angle=-90, width=9cm, clip=true]{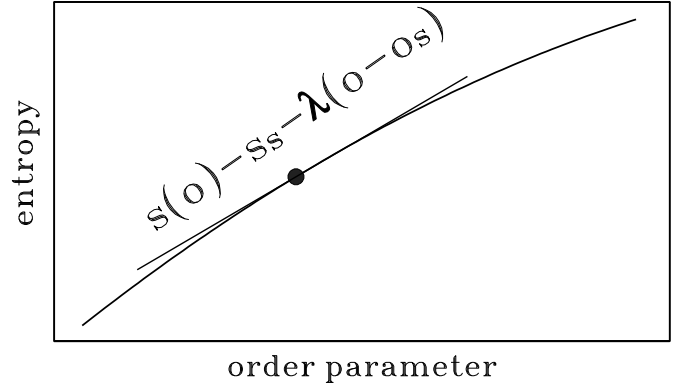}}
\caption{Mono-phase. The order parameter $o$ is defined in subsection 
  ``order parameter''.\label{concave}}
\end{figure}
\item A transition of first order with
  phase-separation\index{phase-separation} and surface
  tension\index{surface tension} is indicated by the maximum curvature
  $\lambda_1(e,n)>0$.  $s(e,n)$ has a convex intruder (upwards bended
  with Pseudo-Riemannian geometry\index{Pseudo-Riemannian geometry})
  in the direction $\vecb{v}_1$ of the largest curvature (order
  parameter).  There are at least three solutions ($e_s,n_s$):
  $o_1,o_2,o_3$ see figure~(\ref{convex}) of
\begin{eqnarray}
\beta=\frac{1}{T}&=&\left.\frac{\partial S}{\partial E}\right|_{e_s,n_s}\\
\nu=-\frac{\mu}{T}&=&\left.\frac{\partial S}{\partial N}\right|_{e_s,n_s}
\end{eqnarray}  
\begin{figure}[h]
\cent{\includegraphics*[bb = 105 15 463 623, angle=-90, width=9cm,
  clip=true]{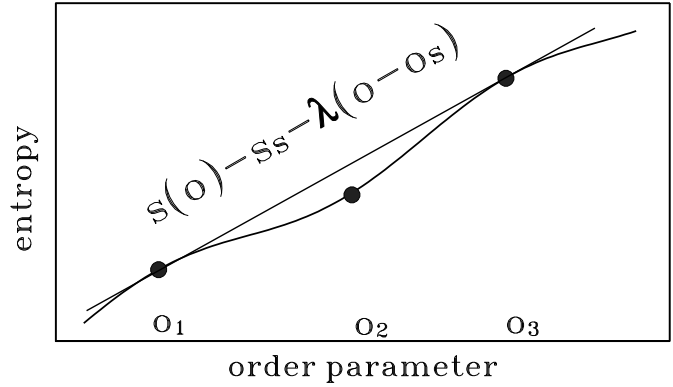}}
\caption{Phase-separation, Gibbs double tangent \label{convex}}
\end{figure}
The whole region \{$o_1,o_3$\} is mapped by eq.~(\ref{grandsum}) into
a single point ($T_{tr},\nu_{tr}$) in the canonical ensemble which is
consequently non-local in $o$ and {\em bimodal}.  \index{bimodality of
  the canonical ensemble at phase transitions} This is the origin of
the\index{Yang-Lee singularity} Yang-Lee singularities. I.e.  if the
largest curvature of $s(e,n)$ is $\lambda_1\ge 0$ both ensembles are
not equivalent as already pointed out by us in
1993\cite{gross124,gross140,gross158,gross174} see also Barr\'{e} et 
al~\cite{barre01}. The
possibility of positive curvatures (Pseudo Riemannian
geometry\index{Pseudo-Riemannian geometry}) is the main difference to
the Riemannian geometry proposed by\index{phase-separation} G.
Ruppeiner\cite{ruppeiner95,andresen96}.  Ruppeiner discusses shortly
phase transitions of first order and points out that there is no
divergence of the correlation length $\xi$ like at second-order
transitions and $\xi$ remains of the order of the interface thickness,
i.e. finite.  He does not mention (or is not interested in) the
positivity of the curvature of $s(e,n)$, which encodes important
information about the surface tension, see below and which is also the
essential reason why the canonical formalism does not apply to
non-extensive systems.
\item A continuous (``second order'') transition with vanishing
  surface tension\index{surface tension}, where two neighboring phases
  become indistinguishable. I.e. where the three stationary solutions
  $o_1,o_2,o_3$ move into one-another. This is at the extremum of
  $\lambda_1$ in the direction of order parameter
  $\vecb{v}_{\lambda=0}\cdot\vecb{$\nabla$}\lambda_1=0$. These are the
  catastrophes of the Laplace transform eq.(\ref{grandsum}) $E\to T$
  and the critical points\index{critical point} of the transition.  If
  it is also on the border line ($\lambda_1=0$) of the first order
  transition (where $\lambda_1<0$), it is the critical end-point of
  the transition. It is an open question whether a line of
  second-order transition is also the locus of the critical end-point
  of a first-order transition in a hidden parameter \cite{gross174}.
\end{itemize}
\subsubsection{Physical origin of positive curvature, the surface
  tension\index{surface tension}\label{surface tension}}
\begin{figure}[h]\cent{
\includegraphics*[bb = 99 57 400 286, angle=-0, width=9cm,  
clip=true]{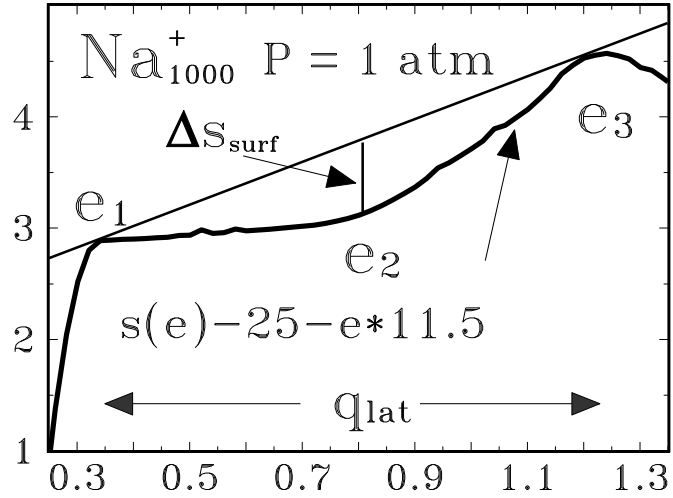}}
\caption{MMMC~\protect\cite{gross174} simulation of the entropy  
  \index{entropy} $s(e)$ per atom ($e$ in eV per atom) of a system of
  $N_0=1000$ sodium atoms at an external pressure of 1 atm.  At the
  energy $e\le e_1$ the system is in the pure liquid phase and at
  $e\ge e_3$ in the pure gas phase, of course with fluctuations. The
  latent heat per atom is $q_{lat}=e_3-e_1$.  \underline{Attention:}
  the curve $s(e)$ is artifically sheared by subtracting a linear
  function $25+e*11.5$ in order to make the convex intruder visible.
  {\em $s(e)$ is always a steeply monotonic rising function}.  We
  clearly see the global concave (downwards bending) nature of $s(e)$
  and its convex intruder. Its depth is the entropy \index{entropy}
  loss due to additional correlations by the
  interfaces\index{entropy-loss due to interface}. It scales $\propto
  N^{-1/3}$. From this one can calculate the surface
  tension\index{surface tension} per surface atom
  $\sigma_{surf}/T_{tr}=\Delta s_{surf}*N_0/N_{surf}$.  The double
  tangent (Gibbs construction) is the concave hull of $s(e)$. Its
  derivative gives the Maxwell line in the caloric curve $T(e)$ at
  $T_{tr}$, fig.~(\ref{entrop2}). In the thermodynamic limit
  \index{thermodynamic limit} the intruder would disappear and $s(e)$
  would approach the double tangent from below.  Nevertheless, even
  there, the probability of configurations with
  phase-separations\index{phase-separation} are suppressed by the
  (infinitesimal small) factor $e^{-N^{2/3}}$ relative to the pure
  phases and the distribution remains {\em strictly bimodal in the
    canonical ensemble}. The region $e_1<e<e_3$ of phase separation
  gets lost.\label{naprl0f}}
\end{figure}

~\\For short-range forces it is linked to the interphase surface
tension.  This is demonstrated for a system of $1000$ Na-clusters by
figure~(\ref{naprl0f}).
\begin{table}
\caption{Parameters of the liquid--gas transition of small
  sodium clusters (MMMC-calculation~\protect\cite{gross174}) in
  comparison with the bulk for a rising number $N_0$ of atoms,
  $N_{surf}$ is the average number of surface atoms (estimated here as
  $\sum{N_{cluster}^{2/3}}$) of all clusters with $N_i\ge2$ together.
  $\sigma/T_{tr}=\Delta s_{surf}*N_0/N_{surf}$ corresponds to the
  surface tension\index{surface tension}. Its bulk value is adjusted
  to agree with the experimental values of the $a_s$ parameter which
  we used in the liquid-drop formula for the binding energies of small
  clusters, c.f.  Brechignac et al.~\protect\cite{brechignac95}, and
  which are used in this calculation~\cite{gross174} for the individual
  clusters.}
\begin{center}
\renewcommand{\arraystretch}{1.4}
\setlength\tabcolsep{5pt}
\begin{tabular} {|c|c|c|c|c|c|} \hline 
&$N_0$&$200$&$1000$&$3000$&\vecb{bulk}\\ 
\hline 
\hline  
&$T_{tr} \;[K]$&$940$&$990$&$1095$&\vecb{1156}\\ \cline{2-6}
&$q_{lat} \;[eV]$&$0.82$&$0.91$&$0.94$&\vecb{0.923}\\ \cline{2-6}
{\bf Na}&$s_{boil}$&$10.1$&$10.7$&$9.9$&\vecb{9.267}\\ \cline{2-6}
&$\Delta s_{surf}$&$0.55$&$0.56$&$0.44$&\\ \cline{2-6}
&$N_{surf}$&$39.94$&$98.53$&$186.6$&$\vecbm{\infty}$\\ \cline{2-6}
&$\sigma/T_{tr}$&$2.75$&$5.68$&$7.07$&\vecb{7.41}\\
\hline
\end{tabular}
\end{center}
\end{table}
\newpage
\subsubsection{Negative heat capacity\index{negative heat capacity} 
  as signal for a phase transition of first
  order\label{negative heat-capacity}}
\begin{figure}[h]
\begin{minipage}[t]{8cm}
\includegraphics*[bb =83 58 379 573, angle=-180, width=8 cm,  
clip=true]{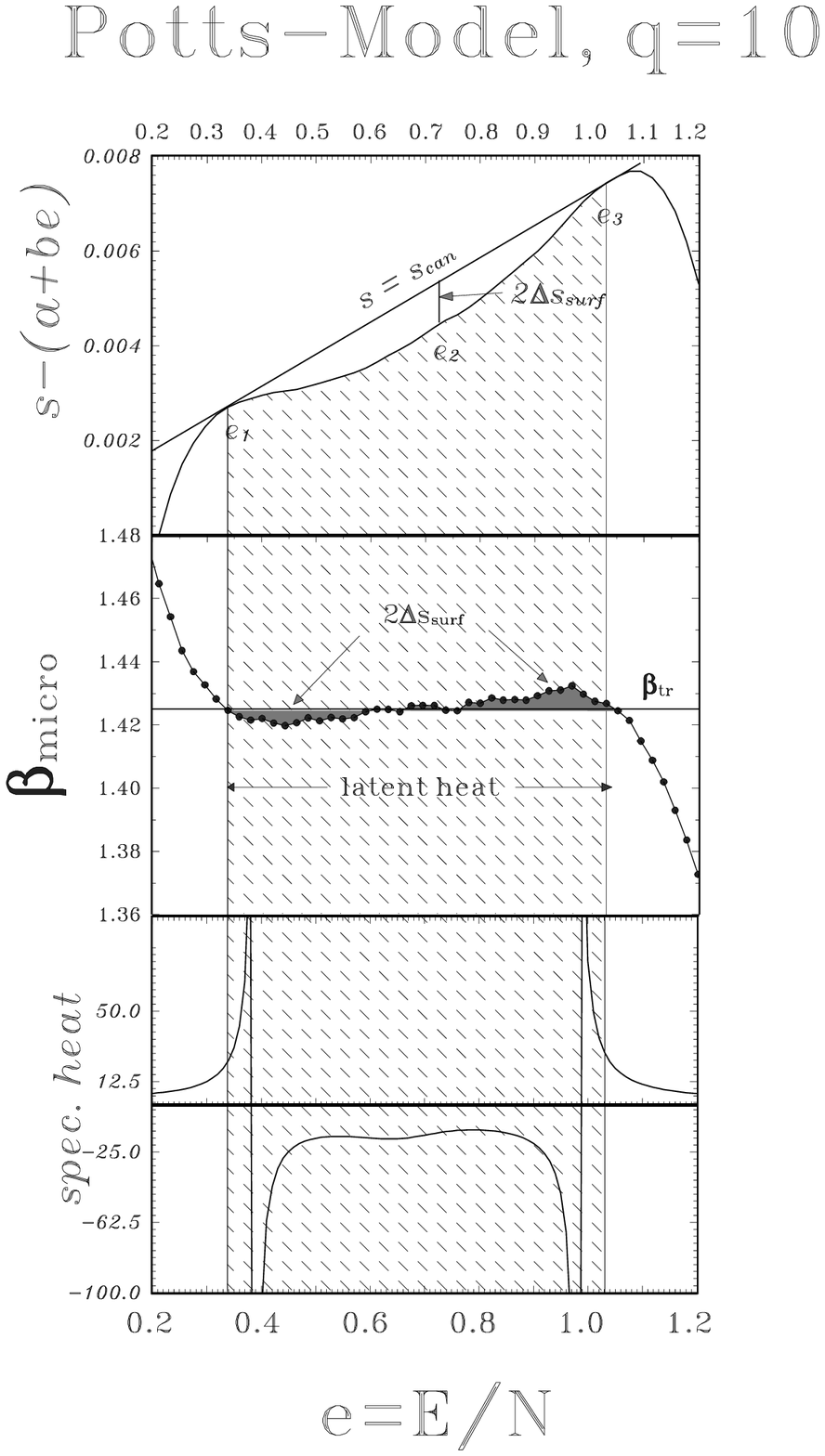}
\end{minipage}
\begin{minipage}[t]{4cm}
\caption{a) Specific entropy
  \protect{$s( e)=\int_0^{ e}{\beta_{micro}(\bar{ e}) d\bar{ e}}$} vs.
  the specific energy $ e$ for the 2-dim.  Potts model with $q=10$
  spin orientations per lattice point on a $100*100$ lattice.  In
  order to visualize the anomaly of the entropy the linear function
  $a+b e$ ($a=s(0.2119)$, $b=1.4185$) was subtracted. Because we use
  periodic boundary conditions one needs two cuts to separate the
  phases and the depth of\index{entropy, convex intruder of} the
  convex intruder is twice the surface-entropy.  \protect\newline b)
  Inverse temperature $\beta_{micro}( e)=1/T( e)$ as directly
  calculated by $M\!M\!M\!C$ \protect\newline
  c) Specific heat $c( e)=-\beta^2/(\partial\beta/\partial e)$.  The
  canonical ensemble of the bulk jumps over the shaded region between
  the vertical lines at $ e_1$ and $ e_3$. This is the region of the
  coexistence of two phases one with ordered spins, the other with
  disordered spins. Here $c( e)$ has two poles and in between it
  becomes negative.  \index{specific heat, negative} Canonical
  thermodynamics is blind to this region. Observe that the poles are
  {\em inside} $ e_1\le e\le e_3$, i.e the canonical specific heat
  (non-dashed region) remains finite and positive as it should, from
\cite{gross150}.}
\label{entrop2}
\end{minipage}
\end{figure}

~\\
As explained in figure~(\ref{entrop2}) for the example of the
$q=10$ Potts-model, a positive curvature (convex intruder) of $S(E)$
in the energy direction corresponds to a characteristic {\em
  backbending} of the caloric curve $T(E)$ or $\beta(E)$, and to a
negative heat-capacity $c$, the general signal for a phase transition of
first order as proposed by us more than $15$ years 
ago~\cite{gross82,gross95,gross124,gross150,gross174,gross176}:
\begin{eqnarray}
c&=&\frac{\partial E}{\partial T}\nonumber\\
&=&-\left(\frac{\partial S}{\partial E}\right)^2/\frac{\partial^2
S}{\partial^2E}
\end{eqnarray} This was later-on further elaborated by Chomaz and 
Gulminelly~\cite{gulminelli99a,gulminelli99,chomaz99,chomaz00a,chomaz00}
and experimentally verified~\cite{schmidt00,dAgostino00}.

Connecting such a system at $e_1 +\delta E,T_1$ with another one at
$e_3-\delta E>e_1+\delta E, T_3$ and $T_1>T_3$ then the latter one
heats up to $T_{combined}=T_3+\Delta T$ by loosing energy, whereas the
former one cools down to $T_{combined}=T_1-\Delta T$ by gaining
energy.  Here one of the Clausius formulation of the Second
Law\index{Second Law} gets invalidated: {\em ``Heat flows always from
  the hotter to the cooler body.''}
\index{Clausius formulation of Second Law violated: 
Heat can flow from cold to hot} Or with other words in the region
of negative heat capacity a system acts {\em in equilibrium} as a
refrigerator. This is well within ordinary classical thermodynamics!
However, here Ruppeiner's Riemannian geometry~\cite{ruppeiner95} must
be extended to a Pseudo-Riemannian geometry. 
\index{Pseudo-Riemannian geometry}
\newpage
\subsubsection{The topology of the curvature $c(e,n)$ de-covers the global
  phase-diagram}
\begin{figure}[h]
\includegraphics*[bb =0 0 290 180, angle=0, width=12cm, 
clip=true]{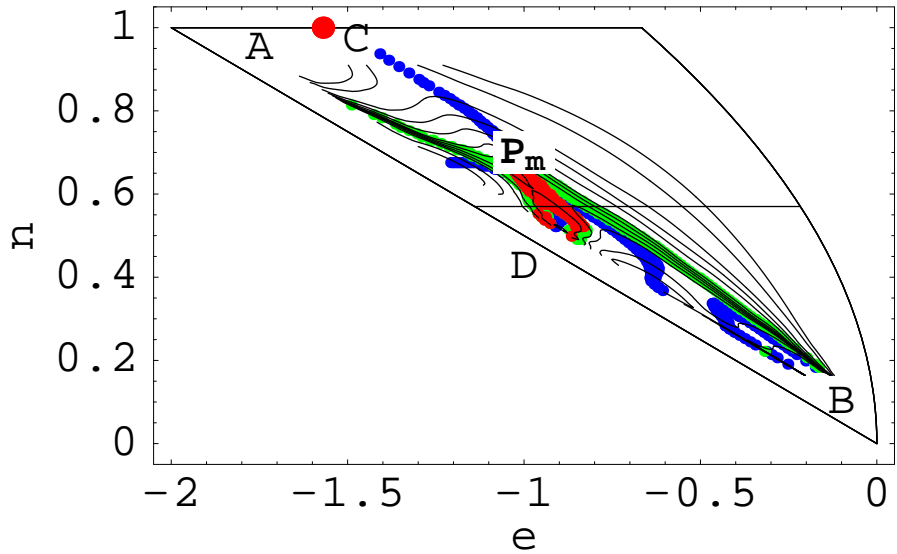}
\caption{Global phase diagram or conture plot of the curvature determinant
  (Hessian), eqn.~(\ref{curvdet}), of the 2-dim Potts-3 lattice gas
  with $50*50$ lattice points, $n$ is the number of particles per
  lattice point, $e$ is the total energy per lattice point;The line
  (-2,1) to (0,0) is the ground-state of the lattice-gas
  eq.~(\ref{hamiltonian}) as function of $n$. The most right curve is
  the locus of configurations with completely random spin-orientations
  (maximum entropy). The whole physics of the model is between these
  two boundaries.  At the green lines the Hessian is $\det=0$,
  this is the boundary of the region of phase
  separation\index{phase-separation} (the triangle $AP_mB$) with a
  negative Hessian.  \index{Hessian of entropy determines
  phase-structure} The region of Pseudo-Riemannian
  geometry\index{Pseudo-Riemannian geometry}; At the blue lines
  is a minimum of $\det(e,n)$ in the direction of the largest
  curvature ($\vecbm{v}_{\lambda_{max}}\cdot\vecbm{\nabla}\det=0$),
  these are lines of second order transition; In the triangle $AP_mC$
  is the pure ordered (solid) phase ($\det>0$); Above and right of the
  line $CP_mB$ is the pure disordered (gas) phase ($\det>0$); The
  crossing $P_m$ of the boundary lines is a multi-critical point. It
  is also the critical end-point of the region of phase separation
  ($\det<0$).  The red region around the multi-critical point
  $P_m$ corresponds to a flat (cylindric) region of $\det(e,n)\sim 0$
  and {\boldmath$\vecbm{\nabla}$} \mbox{$\lambda_1$}{\boldmath$\sim
  0$}, details see \protect\cite{gross173}; $C$ is the analytically
  known position of the critical point which the ordinary $q=3$ Potts
  model (without vacancies){\em would have in the thermodynamic limit}
  \index{thermodynamic limit} $N\to\infty$, from~\cite{gross173}.
  \label{det}}
\end{figure}

\index{global phase diagram} 
~\\ Figure~(\ref{det}) shows the example of a micocanonical global
phase diagram of the Potts ($q=3$) lattice-gas as function of the
energy $e=E/L^2$ per lattice-point and the relative occupancy
$n=N/L^2$~\cite{gross173}.  $L\times L$ is the size of the lattice
taken to be $L=50$, and $0\leq N\leq L^2$ is the number of occupied
sites.

The Hamiltonian of the lattice gas is:
\begin{eqnarray}
H&=&-\sum_{i,j}^{n.n.pairs}o_i o_j\delta_{\sigma_i,\sigma_j}
\label{hamiltonian}\\
n&=&L^{-2}N=L^{-2}\sum_io_i .\nonumber\\
\mbox{with the occupancy }o_i&=&\left\{\begin{array}{cl}
1&\mbox{, spin particle in site }i\\
0&\mbox{, vacancy in site }i\\
\end{array}\right. .\nonumber
\end{eqnarray}
\newpage
\subsubsection{Order parameters}\label{order parameter}
\begin{center}\begin{figure}[h]
    \includegraphics*[bb =0 0 290 180, angle=-0,
    width=11cm,clip=true]{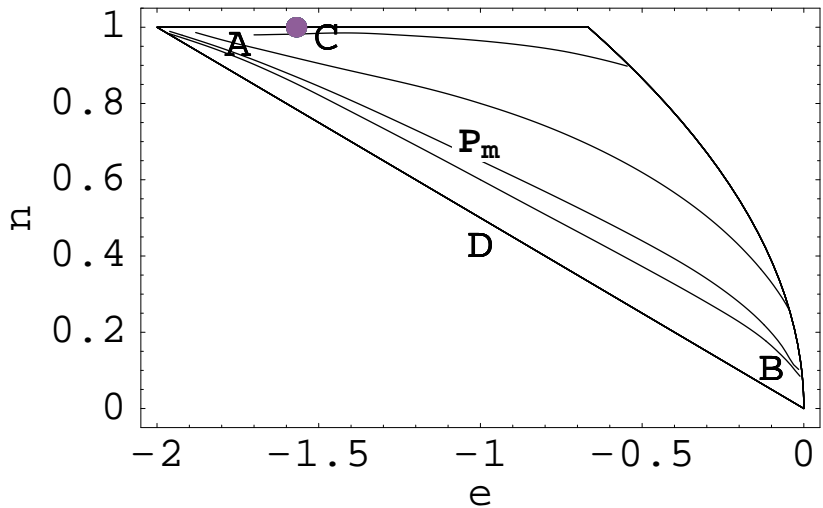}
\caption{Various trajectories of maximum curvature $\lambda_1$,
  $\mbox{\boldmath$v_1$}=$order parameter, for the global
  phase-diagram of fig.~(\ref{det}).  Left and below the
  multi-critical point $P_m$\index{multi-critical point}, in the
  region of phase-separation (positive maximum curvature $\lambda_1$)
  we see an approximately linear behavior. Here $\beta(e,n)=$const.~is
approximately paralell to $\nu(e,n)=$const.~and paralell to 
$\mbox{\boldmath$v_1$}(e,n)$.
\label{maincurvature}}
\end{figure}\end{center}

~\\
Definition: In the geometric theory, the order parameter $o$ of a
phase-transition is defined as the length of the trajectory along the
direction of maximum curvature in the global phase-diagram see
fig~(\ref{maincurvature}).  Progressing in that direction one
experiences the transition from one phase to the other. In the
thermodynamic limit (when it exists) the region of positive maximum
curvature is jumped over by the canonical ensemble and the order
parameter jumps here also (traditional definition of the order
parameter). It is important to notice that the order parameter is {\em
  not} a simple linear function of the control
parameters\index{control parameter} like $E,N$ c.f.
fig.~(\ref{maincurvature}).

If there are more control parameters\index{control parameter}
($e,n,\cdots$) there might be a situation where because of some
underlying symmetry the eigenvalue 
\index{curvature, largest eigenvalue} of largest curvature 
$\lambda_1$ is degenerate. In those cases the order parameter is
multi-dimensional.  All these features, convex regions, curvatures of
$s(e,n,\cdots)$ are {\em topological} properties of the entropy
surface\index{entropy surface} $s(e,n,\cdots)$ determined by the area
of the manifold of constant energy etc. within the N-body phase space.
\vspace*{-0.5cm}
\subsubsection{Ambiguity of intensive parameters and the canonical ensemble}
\label{ambiguity}~\\
\index{multi-valuedness of entropy in terms of intensive variables}
\index{redundancy of intensive parameters}
\begin{figure}[h]\cent{
\includegraphics*[bb = 0 0 290 280, angle=-0, width=9cm,
  clip=true]{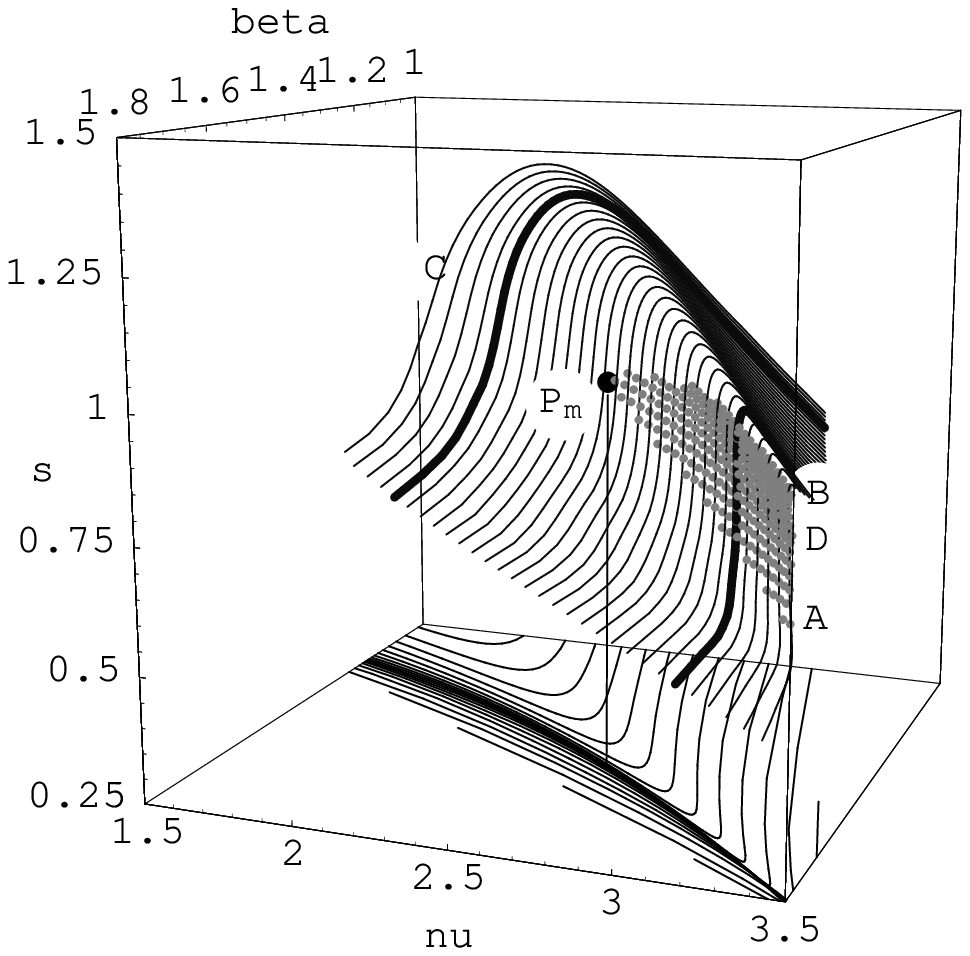}}
\caption{Plot of the entropy\index{entropy} $s(e,n)$ 
  as function $s_{micro}(\beta,\nu)$ of the ``intensive''
  variables\index{intensive variables} ($\nu=\partial s/\partial
  n=-\beta\mu$) in the figure labeled as ``nu'' and $\beta=\partial
  s/\partial e$ is called ``beta''.  The lines which build the surface
  are lines for $\beta=$cons. The positions of the points
  $A$,$D$,$B$,$C$ defined in fig.~(\ref{det}) are only roughly
  indicated. The convex intruder\index{entropy, convex intruder of}
  between the lines $\widehat{AP_mB}$ and the ground-state line
  $\widehat{ADB}$ of fig.~(\ref{det}), where $s_{micro}(\beta,\nu)$
  becomes multi-valued as function of $\nu>\nu_{P_m}$ and
  $\beta>\beta_{P_m}$, here seen from the side, is indicated by
  shadowing.  This corresponds to phase-separation at a first order
  transition. At the bottom the projection of the entropy
  surface\index{entropy surface, projection on intensive parameters} 
  onto the \{$\beta, \nu$\} plane is shown as contour plot (lines of
  equal $S(\beta,\nu)$). The convex part (region of
  phase-separation)\index{phase-separation} is hidden behind the dark
  ``critical'' line in the ($\beta,\nu$)-plane\index{critical line}.
\label{Sintens}}
\end{figure}
Fig.~(\ref{Sintens}) explains what happens if one plots the entropy
$s(e,n)$ vs. the ``intensive'' quantities $\beta=\partial S/\partial
E$ and $\nu =\partial S/\partial N$ as one would do for the
grand-canonical ensemble: As there are several points $e_i,n_i$ with
identical $\beta,\nu$, $s_{micro}(\beta,\nu)$ is a {\em multi-valued
  function of $\beta,\nu$}.\index{entropy surface} The entropy surface
$s_{micro}(e,n)$ is folded onto itself.  In the projection in
fig.~(\ref{Sintens}) on the $\beta,\nu$-plane, these points show up as
a black critical line (dense region). This black line continues over
the multi-critical point $P_m$ towards $C$ indicating the direction
towards the critical point of the ordinary $q=3$ Potts model at $n=1$
(zero vacancies). Between $P_m$ and $C$ the slopes
\begin{eqnarray}
\left.\frac{\partial s}{\partial \beta}\right|_\nu&=&
\frac{1}{d}[\beta s_{nn}-\nu s_{ne}]\\
\mbox{or}&&\\
-\left.\frac{\partial s}{\partial \nu}\right|_\beta&=&
\frac{1}{d}[\beta s_{en}-\nu s_{ee}]
\end{eqnarray}
are negative large but finite.

The information given by the projection would be all information which
can be obtained from the conventional grand-canonical
entropy\index{entropy, grand canonical} $s(T,\mu,V)$, if we would have
calculated it from the Laplace transform, eq.(\ref{grandsum}).

The back folded branches, the convex intruder 
\index{entropy, convex intruder of} of $s(e,n)$, the region of
phase-separation,\index{phase-separation} is jumped over in the
Laplace transform eq.  (\ref{grandsum}) and gets consequently lost in
$Z(T,\mu)$. Here $s(T,\mu)$ becomes non-analytical \lora Yang-Lee
singularity. This demonstrates the far more detailed insight into
phase transitions and critical phenomena obtainable by the geometrical
interpretation of microcanonical thermo-statistics~\cite{gross186}
but which is not accessible by the canonical treatment, c.f. the similar
arguments of Gibbs \cite{gibbs06}.
\subsection{Rotating self-gravitating systems\index{gravitation}}\label{gravit}
\index{long range interaction}\index{non-extensivity}\index{rotation}
\index{angular momentum and gravity}
\subsubsection{Stars and multi-star clusters}~\\
The most interesting and important non-extensive systems are
self-gravitating ones. I.e. systems with the Hamiltonian
\begin{equation} H_N\equiv
H_N(\{\vecbm{r}_i\},\{\vecbm{p}_i\})=\frac{1}{2m}
\sum_{i=1}^N p_i^2+\Phi(\{\vecbm{r}_i\})
\end{equation}
with the gravitation $\Phi(\{\vecbm{r}_i\})=-Gm^2\sum_{i<j}
|\vecbm{r}_i-\vecbm{r}_j|^{-1}$. $\vecbm{r}_i$,
$\vecbm{p}_i$ and $m$ denote, respectively, the position,
momentum and mass of the $i$-th particle.  Because of the long range
of the gravitation the total potential energy is $\propto N^2$ and
consequently non-extensive.

The statistical equilibrium of self-gravitating systems without
angular-momen\-tum was first considered by Thirring~\cite{thirring70}
who pointed out that microcanoni\-cal\-ly these systems have a
negative heat capacity\index{negative heat capacity} and therefore the
microcanonical and canonical ensembles are not
equivalent\index{non-equivalence of ensembles}.

Here, we overcome the simple Thirring model and investigate the
equilibrium of rotating, self-gravitating systems in a spherical
symmetric box under various energies $E$ and angular-momenta $L$.
These calculations are done by E.V.Votyakov~\cite{gross187}.
\\The following approximations are used:\\
a)~Mean-field approximation: We approximate any $N$-body spatial
density $\rho_N=\rho(\vecbm{r}_1,\dots,\vecbm{r} _N)$ by the
non-correlated product of single-particle densities~\cite{gross187}
and work henceforth in the single-particle $\mu$-space.
\index{mean-field approximation}
\begin{eqnarray}
\rho_N(\vecbm{r}_1,\dots, \vecbm{r}_N)&\approx&\prod_{i=1}^N
{\rho(\vecbm{r}_i)}\\
\mbox{and the gravitation interaction:}\nonumber\\
\Phi[\rho_N]&\approx&
-Gm^2\sum_{i<k}\int{\frac{\rho(\vecbm{r}_i)\rho(\vecbm{r}_k)}
{|\vecbm{r}_i-\vecbm{r}_k|}d\vecbm{r}_id\vecbm{r}_k}
\end{eqnarray}

The aim is now to find those density profiles $\rho(\vecbm{r})$ that
maximize the (mean-field) entropy ($k=1$)
\begin{equation}\label{duino}
S^m_N(E,\vecbm{L})=\ln W^m_N(E,\vecbm{L})
\end{equation}
$W^m_N$ being the mean-field approximation to the
microcanonical ``partition sum'', i.e. the sum of
all uncorrelated many-body states ($W^m_N\le W_N$).
{\em After integrating over the $N$-momenta}:
\begin{equation}\label{wu3}
W^m_N(E,\vecbm{L})\!=\!\frac{A}{N!}\int\!\l[E-\frac{1}{2}\vecbm{L}^T
\mathbb{I}^{-1}\vecbm{L}-\Phi[\rho]\r]^{\frac{3N-5}{2}}
\!\!\!\!P[\rho]d\rho(\vecbm{r})
\end{equation}
where $P[\rho]$ is the probability to observe a density profile
$\rho\equiv \rho(\vecbm{r})$.

We use the same trick as Lynden-Bell~\cite{lyndenbell67} and avoid
configurations with high densities where other physical processes like
nuclear reactions become more important than gravity. To achieve this,
we subdivide the spherical volume $V$ into $K$ identical cells labeled
by the positions of their centers.  The idea is to replace the
integral over $V$ with a sum over the cells. In order to avoid
overlapping and to cure the short-distance singularity of the
Newtonian potential we assume that each cell may host up to $n_0$
particles ($1\ll n_0\ll N$ but $Kn_0>N$). This condition is
essentially equivalent to consider hard spheres instead of point
particles. The probability $P[\rho]$ to find a given density profile
$\rho(\vecbm{r})$ is now proportional to the number of ways in which
our $N$ particles can be distributed inside the $K$ cells with maximal
capacity $n_0$ and individual occupancies $n(\vecbm{r}_k)$ or
$\rho(\vecbm{r}_k)=n(\vecbm{r}_k)K/V$. A simple combinatorial reasoning
leads to
\begin{equation}
P(\rho)=\prod_{{\rm cells~} k,\sum{n(r_k)}=N}\binom
{n_0}{n(\vecbm{r}_k)}.
\end{equation}
This looks analogous to a Fermi-Dirac statistics, however, here {\em
  only in coordinate space}.  
\index{Fermi-Dirac statistics in position} This guarantees the strict
non-overlapping condition. $\rho$ can nowhere be larger than $\rho_0=n_0K/V$.
In contrast, in a recent paper Chavanis and
Ispolatov~\cite{chavanis02a} use Fermi-Dirac statistics in {\em
phase-space}.

\begin{figure}[h]\begin{center}
    \includegraphics*[bb =0 540 408 809, angle=-0,
    width=11cm,clip=true]{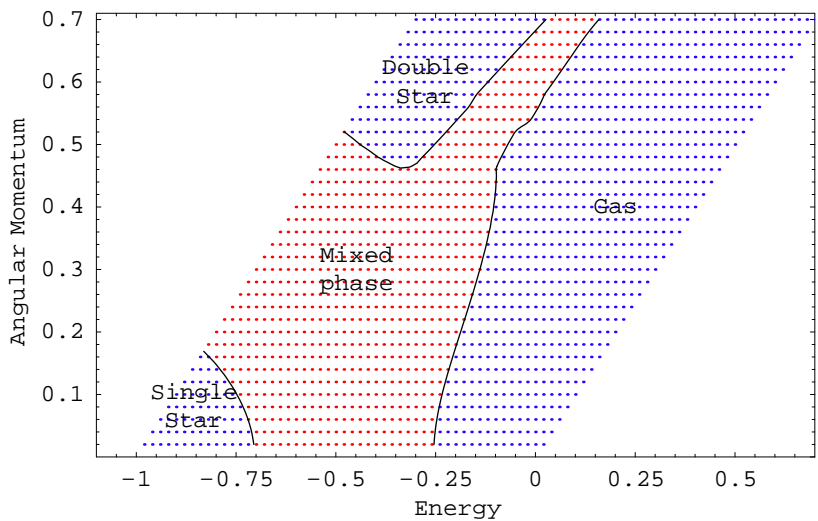}
\caption{Global phase diagram of a self-gravitating system as
  function of energy $E$ and angular-momentum $L$ in dimension-less
  units.  \index{global phase diagram} Systematic calculations were
  done only in the dotted region. There are 3 different mono-phases
  (regions of negative maximum curvature $\lambda_1<0$): A single star
  plus some gas at small $E$ and $L$; at high energy and moderate $L$,
  a gas phase; and at large $L$ configurations of broken spherical
  symmetries with double-star systems. The mixed phase is the region
  with positive maximum curvature $\lambda_1>0$ where at least two
  different phases (single star and gas (low $L$), single star and
  double star systems, eventually rings, at intermediate $L$ are in
  competition, finally double-stars and gas compete with one-another
  at large $L$ and $E$.\label{globphd}\index{gravitation}}
\end{center}\end{figure}

We express $S^m_N$ as a functional of the density profile
$\rho(\vecbm{r})$, such that
\begin{equation}
\int_V\rho(\vecbm{r})d\vecbm{r}=N
\end{equation}
and subsequently find the $\rho(\vecbm{r})$'s that maximize $S^m_N$.
This leads to the self-consistent integral equation:
\begin{equation}\label{key}
\log\frac{\rho(\vecbm{x})}{\rho_0-\rho(\vecbm{x})}=
-\frac{\beta}{\Theta}U(\vecbm{x})+\frac{1}{2}\beta(
\boldsymbol{\omega}\times\vecbm{x})^2-\mu
\end{equation} 
or, equivalently,
\begin{equation}\label{key2}
\rho(\boldsymbol{x})=\frac{\rho_0}{1+e^{\frac{\beta}{\Theta}U(\vecbm{x})
-\frac{1}{2}\beta(\vecbm{\omega}\times\vecbm{x})^2+\mu}}
\end{equation}
where $\vecbm{\omega}[\rho]$ is the angular velocity (related to
the total angular momentum by the relation
$\vecbm{L}=\vecbm{I}[\rho]\vecbm{\omega}[\rho]$), and $\beta[\rho]$ and
$U(\vecbm{x})$ are respectively defined as
\begin{gather}
\beta=\frac{3/2}{E-\frac{1}{2}\vecbm{L}^T
(\vecbm{I}[\rho])^{-1}\vecbm{L}-\Phi[\rho]}\\
U(\vecbm{x})=-\int\frac{\rho(\vecbm{x'})}{
|\vecbm{x}-\vecbm{x'}|}~d\vecbm{x'}\label{Uofx}\\
\Phi[\rho]=\frac{1}{2}\int{U(\vecbm{x})\rho(\vecbm{x})d\vecbm{x}}
\end{gather}
b)~To solve eq.(\ref{key2}), Votyakov~\cite{gross187} expands 
$\rho(\vecbm{r}_i)$ into spherical harmonics:
\begin{equation}
\rho(\vecbm{r}_i)=\sum_{l,m}{b_{l,m}(|r_i|)\;Y_{lm}(\Theta_i,\phi_i)}
\end{equation}
and ignores for simplicity reasons all odd $l,m$ as well also all
$l>16$. I.e. he allows only for parity even, upside -- down symmetric
configurations (this is later overcome). 
The expansion into spherical harmonics has the
advantage that the original non-linear self-consistent
three-dimensional integral equation (\ref{key2}) becomes now a finite
self-consistent set of coupled one-dimensional and two-dimensional
integral equations. 
As function of energy and total angular-momentum the
microcanonical global phase-diagram defined by the topology of the
curvature (Hessian) of $S(E,N)$ shows an astonishing rich picture see
fig.(\ref{globphd}).

This example proves the superiority of the geometrical, topological,
interpretation of thermo-statistics. It reproduces realistically many
different configurations of even gravitating systems. No canonical
description nor Tsallis non-extensive statistics had achieved this.
\newpage
\section{Approach to equilibrium, Second Law}
\index{Second Law}
\subsection{Zermelo's paradox}\index{Zermelo's paradox}
When Zermelo~\cite{zermelo97} argued against Boltzmann, that following
Poncarr\'e any many-body system must return after the Poincarr\'e
recurrence time\index{Poincarr\'e recurrence time} $t_{rec}$ and
consequently its entropy cannot grow indefinitely, 
Boltzmann~\cite{boltzmann1896} answered that for any macroscopic
system $t_{rec}$ is of several orders of magnitude larger than the age
of the universe, c.f.  Gallavotti~\cite{gallavotti99}. Still today,
this is the answer given when the Second Law is to be proven
microscopically, c.f.~\cite{gilbert00}. Then, Zermelo's paradox
becomes blunted.

Here, I argue, even a {\em small} system approaches equilibrium with a
rise of its entropy $\Delta S\ge 0$ under quite general conditions.
Thus, Zermelo's objection must be considered much more seriously.

However, care must be taken, Boltzmann's definition of entropy
eq.(\ref{boltzmentr1}) is only for systems at equilibrium.  To be
precise: in the following I will consider the equilibrium manifold
${\cal{E}}(E,V_a)$ at $t\le t_0$.  At $t_0$ the macroscopic constraint
$V_a$ is quickly removed e.g.  a piston pulled quickly out to
$V_a+V_b$, and the ensemble is followed in phase-space how it
approaches the new equilibrium manifold ${\cal{E}}(E,V_a+V_b)$ see
fig.(\ref{spaghetti}).
\subsection{The solution}
Entropy does not refer to a single point in $N$-body phase space but
to the whole ensemble ${\cal{E}}$ of points.  It is the $\ln(W)$ of
the geometrical size $W$ of the ensemble. Every trajectory starting at
different points in the initial manifold
${\cal{M}}(t=t_0)={\cal{E}}(E,V_a)\in{\cal{E}}(E,V_a+V_b)$ spreads
in a non-crossing manner over the available phase-space
${\cal{E}}(E,V_a+V_b)$ but returns after $t_{rec}$.  Different points
of the manifold, or trajectories, have different $t_{rec}$ which are
normally incommensurable. I.e. the manifold ${\cal{M}}(t)$ spreads
irreversibly over ${\cal{E}}(E,V_a+V_b)$.
\subsubsection{Mixing}\index{mixing}~\\
When the system is dynamically mixing then the manifold ${\cal{M}}(t)$
will ``fill'' the new ensemble ${\cal{E}}(E,V_a+V_b)$. Though at
finite times the manifold remains compact due to Liouville and keeps
the volume $W(E,V_a)$, but as already argued by
Gibbs~\cite{gibbs02,gibbs36} ${\cal{M}}(t)$ will be filamented like
ink in water and will approach any point of ${\cal{E}}(E,V_a+V_b)$
arbitrarily close. Then, $\lim_{t\to\infty}{\cal{M}}(t)$ becomes dense
in the new, larger ${\cal{E}}(E,V_a+V_b)$. The {\em closure}
$\overline{{\cal{M}}(t=\infty)}$ becomes equal to
${\cal{E}}(E,V_a+V_b)$. I.e. the entropy $S(t=\infty)>S(t_0)$. This is
the Second Law for a {\em finite} system.
\subsubsection{Macroscopic resolution, fractal distributions and
  closure~\cite{gross183}}
\begin{figure}[h]
\begin{minipage}[h]{5.7cm}
\begin{center}$V_a$\hspace{2cm}$V_b$\end{center}
\vspace*{0.2cm} \includegraphics*[bb = 0 0 404 404, angle=-0,
width=5.7cm, clip=true]{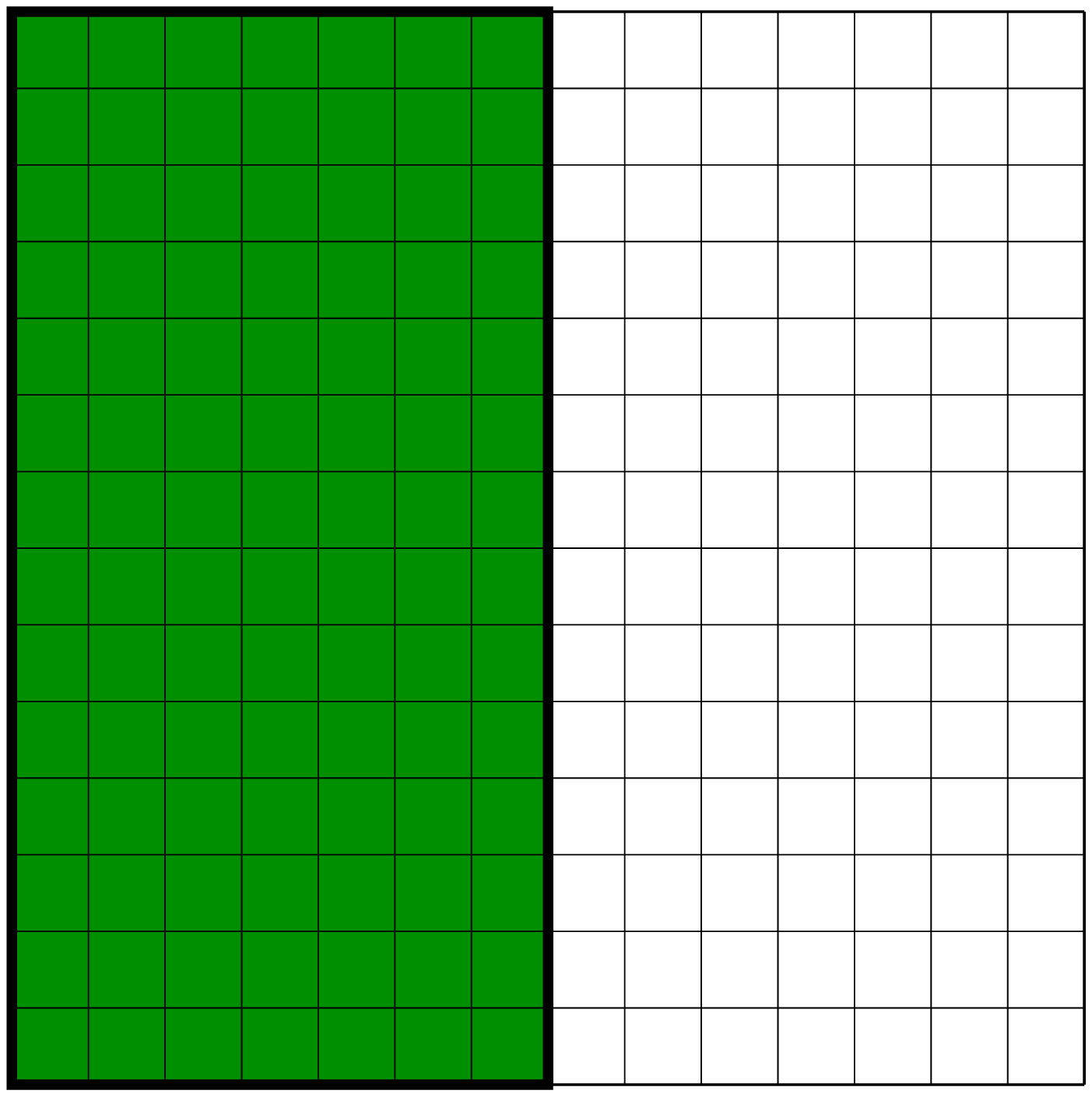}\begin{center}$t<t_0$\end{center}
\end{minipage}\lora\begin{minipage}[h]{5.7cm}
\begin{center}$V_a+V_b$\end{center}\vspace{-0.8cm}
\includegraphics*[bb = 0 0 490 481, angle=-0, width=6.9cm,
clip=true]{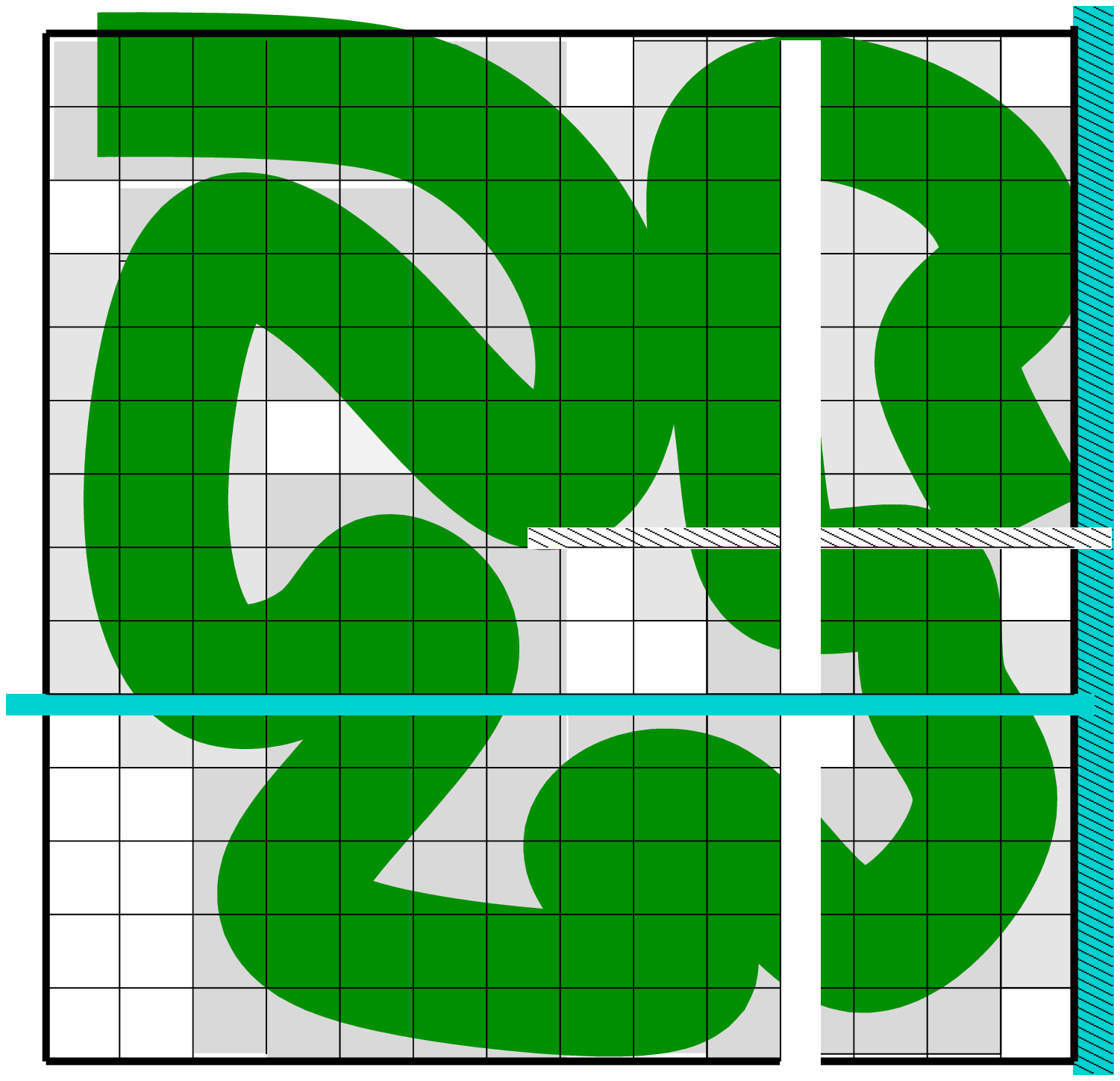}\begin{center}$t>t_0$\end{center}
\end{minipage}\\
\caption{The compact set ${\cal{E}}(t_0)$, left side, develops into an
  increasingly folded but non-crossing ``spaghetti''-like distribution
  ${\cal{E}}(t,t_0)$ in the phase-space with rising time $t$ after
  opening the volume $V_b$. The right figure shows only the early form
  of the distribution. At much later times it will become more and
  more fractal and finally dense in the new larger phase space.  The
  grid illustrates the boxes\index{box-counting} of the box-counting
  method.  All boxes which overlap with ${\cal{M}}(t,t_0)$ contribute
  to the box-counting volume and are shaded gray.  Their number is
  $N_\delta$ \label{spaghetti}}
\end{figure}

~\\
\index{closure of manifold} \index{fractal} We calculate the closure
of the ensemble by box counting~\cite{falconer90}.  Here the
phase-space is divided in $N_\delta$ equal boxes of volume
$\delta^{6N}$. The number of boxes which overlap with ${\cal{M}}(t)$
is $N_\delta$ and the box-counting volume is
then:
\begin{eqnarray}
\Omega_d(\delta)&=&N_\delta \delta^d\mbox{,\hspace{1cm}here with $d=6N-1$}\\
\Omega_d&=&\lim_{\delta\to 0}\Omega_d(\delta).
\end{eqnarray}
The box-counting method is illustrated in fig.(\ref{spaghetti}).  The
important aspect of the box-counting volume of a manifold is that it
is equal to the volume of its closure.

At finite times ${\cal{M}}(t)$ is compact. Its volume $W(t)$ equals that of
its closure $\equiv W(t_0)$. However,
calculated with finite resolution $\delta>0$, $W_{\delta}(t)$
becomes $\ge W(t)$ for $t$ larger than some $t_\delta$, where 
\begin{eqnarray}
W_{\delta}(t)=&\displaystyle{B_d^\delta
\hspace{-0.5 cm}\int}&{\;\;\frac{d^{3N}p\;d^{3N}q}{(2\pi\hbar)^{3N}}
\;\delta\left(E-H_N\{q(t),p(t);[q(t_0),p(t_0)\in {\cal{E}}(V_a)]\}\right)}\\
\displaystyle{B_d^\delta\hspace{-0.5 cm}\int}{\;\;f(q)d^dq}&=&\Omega_d(\delta)
\overline{f(q)}\nonumber.
\end{eqnarray}
A natural finite resolution would be 
\begin{equation}
\delta=\sqrt{2\pi\hbar}.
\end{equation}
Of course the actual problem will often allow a much coarser resolution
because of the insensitivity of the usual macroscopic observables. Then
the equilibration time $t_\delta$ will also be much shorter.

Thus the new definition of Boltzmann's principle
eqs.(\ref{wenv}) is:
\begin{eqnarray}
S&=&\ln(W_\delta),\\
\lefteqn{\mbox{\hspace{-3cm}or mathematically correct, though unphysical, 
at infinite times:}}
\nonumber\\
S&=&\lim_{\delta\to 0}\lim_{t\to\infty}{\ln(W_\delta(t))}=N\ln[(V_a+V_b)/V_a]
+S(t_0).
\end{eqnarray}

\section{Conclusion}

The geometric interpretation of classical equilibrium Statistical
Mechanics by Boltzmann's principle offers an extension also to the
equilibrium of non-extensive systems.

Because microcanonical Thermodynamics as a macroscopic theory controls
the system by a few, usually conserved, macroscopic parameters like
energy, particle number, etc. it is an intrinsically probabilistic
theory. It describes all systems with the same control-parameters
simultaneously. If we take this seriously and avoid the so called
thermodynamic limit ($\lim_{V\to\infty, N/V=\rho}$), the theory can be
applied to the really large, usually {\em inhomogeneous},
self-gravitating systems.  In chapter (\ref{gravit}) it is shown how
this new approach enables to view many realistic astro-physical
configurations as {\em equilibrium configurations under the control of
  total energy and angular-momentum}, c.f.\cite{gross187}.

Within the new, extended, formalism several principles of traditional
Statistical Mechanics turn out to be violated and obsolete. E.g. we
saw that at phase-separation heat (energy) can flow from cold to hot.
Or phase-transitions can be classified unambiguously in astonishingly
small systems. These are by no way exotic and wrong conclusions. On
the contrary, many experiments have shown their validity.  I believe
this approach gives a much deeper insight into the way how many-body
systems organize themselves than any canonical statistics is able to.
The thermodynamic limit clouds the most interesting region of
Thermodynamics, the region of inhomogeneous phase-separation.

Because of the only {\em one} underlying axiom, Boltzmann's principle
eq.(\ref{boltzmentr1}), the geometric interpretation keeps statistics
most close to Mechanics and, therefore, is most transparent. The Second
Law ($\Delta S\ge 0$) is shown to be valid in {\em closed, small} systems
under quite general dynamical conditions.
\section{Acknowledgement}
I have to thank my various collaborators over the last $10$ years.
Their numerical work produced much of the material presented here.
However, first of all I must mention E.V.Votyakov who developed the
sophisticated program for the Potts lattice-gas and the
self-gravitating and rotating system discussed in section
(\ref{gravit}).


%
\clearpage
\addcontentsline{toc}{section}{Index}
\flushbottom
\printindex

\end{document}